\documentclass[12pt,preprint]{aastex}

\title{OB Stars in the Solar Neighborhood I: Analysis of their Spatial Distribution}

\author{F. Elias}
\affil{Facultad de F\'{\i}sica. Departamento de F\'{\i}sica At\'omica, Molecular y Nuclear. Universidad de Sevilla, Apartado 1065, Sevilla, Spain.}

\author{J. Cabrera-Ca\~no}
\affil{Facultad de F\'{\i}sica. Departamento de F\'{\i}sica At\'omica, Molecular y Nuclear. Universidad de Sevilla, Apartado 1065, Sevilla, Spain.}
\affil{Instituto de Astrof\'{\i}sica de Andaluc\'{\i}a, CSIC, Apartado
3004 Granada, Spain.}

\author{E.J. Alfaro}
\affil{Instituto de Astrof\'{\i}sica de Andaluc\'{\i}a, CSIC, Apartado
3004 Granada, Spain.}

\begin{document}

\begin{abstract}

We present a newly-developed, three-dimensional
spatial classification method, designed to analyze the spatial
distribution of early type stars within the 1 kpc sphere around the
Sun. We propose a distribution model formed by two intersecting disks
-the Gould Belt (GB) and the Local Galactic Disk (LGD)- defined by
their fundamental geometric parameters. Then, using a sample of about
550 stars of spectral types earlier than B6 and luminosity classes
between III and V, with precise photometric distances of less than 1
kpc, we estimate for some spectral groups the parameters of our model,
as well as single membership probabilities of GB and LGD stars, thus
drawing a picture of the spatial distribution of young stars in the
vicinity of the Sun.

\end{abstract}


\section{Introduction}

With the naked eye it is possible to observe that the brightest stars
in the sky are mainly distributed along two great circles forming an
angle of about $20\degr$ between them: the Milky Way and a tilted
strip known as the Gould Belt \citep{h3,s2,g3}. Later studies found
out that the GB is better described as a planar distribution of O and
B stars in the solar neighborhood, inclined with respect to the
Galactic plane \citep{l1,s1}.

Although this structure doesn't show an uniform stellar distribution
(the bulk of the stars tend to form aggregates around the regions of
Orion and Sco-Cen, among others), the fact that the kinematic
behavior of the GB members is different from that of the LGD stars
of the same spectral types \citep{l1,s1}, and that several features
of the local interstellar medium such as dust \citep{g1}, neutral
hydrogen \citep{l2,l3} or molecular clouds \citep{d1} seem to be
associated with the system of OB stars, we can assume that we are
witnessing a star-forming process with a spatial scale length
of 1 kpc. Thus, the concept of a star formation complex proposed by
\citet{e1,e2} and recently revised by \citet{e3} appears to have in the
GB its closest example.

Extensive reviews covering the history of research about the GB and
describing the present state of our knowledge and understanding of
this structure can be found in \citet{p2} and \citet{g4}.

\subsection{A review of classification methods in literature}
When it comes to studying the stellar component of this complex, the
first and most serious problem that arises is to isolate the GB
members from the stars belonging to the LGD field. This problem
has been addressed before by many authors, the proposed methods
being intimately related to their {\it a priori\/} hypotheses about the
geometry of the system (usually, either a real belt or toroid, or a
disk).

The first and most intuitive way of facing the problem that we came
across in scientific literature is, once the position of the stars in
a three-dimensional frame (normally, Cartesian Galactic coordinates
$X$,$Y$,$Z$) is known, then  all or some of the three possible projections
(i.e., $Y$ vs. $X$, $Z$ vs. $X$ and $Z$ vs.  $Y$) can be  plotted, in
order to choose limit distance criteria according to the apparent
position of the GB in each of the coordinate planes. This commonly
translates into defining a "box" for each projection inside which
every star belongs to the GB.  Examples of this procedure can be found
in \citet{p1}, \citet{w1}, \citet{l4} and \citet{m1} among others. The
greatest advantage of this method is its simplicity, but undoubtedly
there is an important contamination of LGD stars in the final
selection of GB members, specially in the diffuse zone of intersection
between the two systems.

A second method is based on maximum likelihood analysis of the star
density projected over the celestial sphere. Assuming that the stars
of both the GB and the LGD are confined around two great circles in
the Galactic latitude vs. longitude projection, the proposed stellar
density of both structures decreases as an exponential function of
the angular distance to those great circles. The analysis of those
distributions provides information about the structure of the GB and
the LGD without having to assign individual membership probabilities
to the stars. Fine examples of this method are seen in \citet{c4},
\citet{t1} or \citet{f1}.

Those two lines of attack have in common in that the classification is
based on strictly spatial criteria, obtained from two-dimensional
projections. But, while the first yields only a raw approximation to
the GB members, the second allows to estimate some structural
parameters of the system, such as its inclination with respect to
the Galactic plane, the longitude of its ascending node, its angular
thickness and the fraction of stars belonging to each group. On the
other hand, this model is based on two rather hard structural hypotheses:
\begin{enumerate}
\item The Sun is located in the center of the GB
\item The GB stars are distributed along a toroidal geometry
\end{enumerate}
Such restrictions impose a limitation on other possible three-dimensional
scenarios (such a disk-like structure), that would be difficult to
adjust within these premises.

Therefore, providing that we know the distances of the stars, the need for
a three-dimensional analysis arises. Here, we can take advantage of all
the spatial information at our disposal and to expand the number of
possible structural scenarios. The first to propose such an approach
to the problem were \citet{s1}. They assumed that both the GB and the
LGD could be represented by two crossed planes around which the stars
are distributed by a law decreasing with the distance to each mid
plane, the parameters that define them being estimated by least-square
fitting. Then, they assign individual membership to the stars
according to their vertical distance $Z$ (in Cartesian Galactic
coordinates) to each plane; i.e. a star belongs to the GB if its $Z$
distance to the GB mid-plane is smaller than that to the LGD mid-plane,
and vice-versa. All this procedure is nested within an iterative
algorithm that re-calculates the equation of the planes and re-assigns
memberships until the convergence is reached. The  disadvantages of this
method are that it produces "artificially sharp surfaces on the
systems on the sides that face each other" \citep{s1}, and that, as it
seems unavoidable for any separation method based only in the spatial
distribution of the stars, in the region of intersection between the
planes the discrimination cannot be fully trusted.

\subsection{Objectives}

We have developed a new three-dimensional spatial classification
method which allows us to estimate the mean planes that define the
GB and the LGD, and the single star probabilities of belonging to
either of them. Essentially, as in \citet{s1}, we obtain the mean
planes by least-squares fitting, but instead of simply assigning
membership by a $Z$-coordinate criterion, we define a
parametric stellar density distribution for each plane that makes it
possible to work with membership probabilities. An essential contribution
is the introduction, for the first time in this kind of studies,
of the detection of outliers in the sample, and the analysis of their
effect on the estimation of the model parameters.

In this article, we shall give a detailed description of our
classification method (Sections 2.1 and 2.2), then we shall test its
validity and limitations using synthetic samples (Section 2.3). A
real sample of OB stars from the {\it Hipparcos\/} Catalogue
\citep{e4}, with precise photometric distances will be compiled in
Section 3, and then we shall apply our separation algorithm to this
sample in order to obtain the structural parameters that best describe
the GB and the LGD (Section 4). Finally, we shall develop a correction
of completeness to enhance our results (Section 5), and we shall resume
the most important conclusions of this article in Section 6.

\section{Classification method}
\subsection{The spatial model}

The most restricting simplification that we must assume to build our
model is that the stars belonging to the GB and the LGD are
concentrated along two planes, the distance $d$ of their members to
the mean planes being distributed according to a parametric
probability density function (pdf). Two types of functions have been
tested in this study: exponential and Gaussian pdfs.

\begin{equation}
  \phi(d) = \frac{1}{h} e^{-\frac{d}{h}} ~;~~~ \phi(d) = \frac{1}{h} e^{-(\frac{d^2}{2 h^2})}
\end{equation}

\noindent where $h$ is, respectively, the scale height and the half-width. An
exponential pdf, aside from being very practical in terms of
calculus, is commonly used for describing the vertical distribution
of the stars in models of galactic disks \citep{b1,g2,c2}. However,
it has the annoying characteristic of having a non-continuous
derivative at its maximum. Thus, an apparently more realistic
function such as a Gaussian pdf has been tested too.

Using heliocentric Galactic rectangular coordinates ($X$,$Y$,$Z$),
where $X$ is positive in the direction of the Galactic center, $Y$
in the direction of Galactic rotation and $Z$ perpendicular to the
Galactic plane so that they form a direct frame, the mean planes may
be expressed by the standard Cartesian plane equation

\begin{equation}
a_1 X + a_2 Y + a_3 Z + a_4 = 0
\end{equation}

Then, the distance of any point ($x$,$y$,$z$) to the plane is simply

\begin{equation}
d = \frac{|a_1 x + a_2 y + a_3 z + a_4|}{\sqrt{a_1^2 + a_2^2 +a_3^2}}.
\end{equation}

Although this definition is useful in computational terms, a better
geometrical understanding is provided by the following parameters:
the inclination with respect to the Galactic plane ($i$), the
Galactic longitude of the ascending node ($\Omega$) and the vertical
distance to the Sun ($Z_0$), where

\begin{equation}
\begin{array}{l}
a_1 = \sin i \sin \Omega\\
a_2 = -\sin i \cos \Omega\\
a_3 = \cos i\\
a_4 = -a_3 Z_0
\end{array}
\end{equation}
\noindent
if we normalize the $a_i$ to $\sqrt{a_1^2 + a_2^2 +a_3^2}=1$ (see
Figure 1 for a visualization of the geometrical scheme of the problem).

\begin{figure}[b]
\epsscale{0.8} \plotone{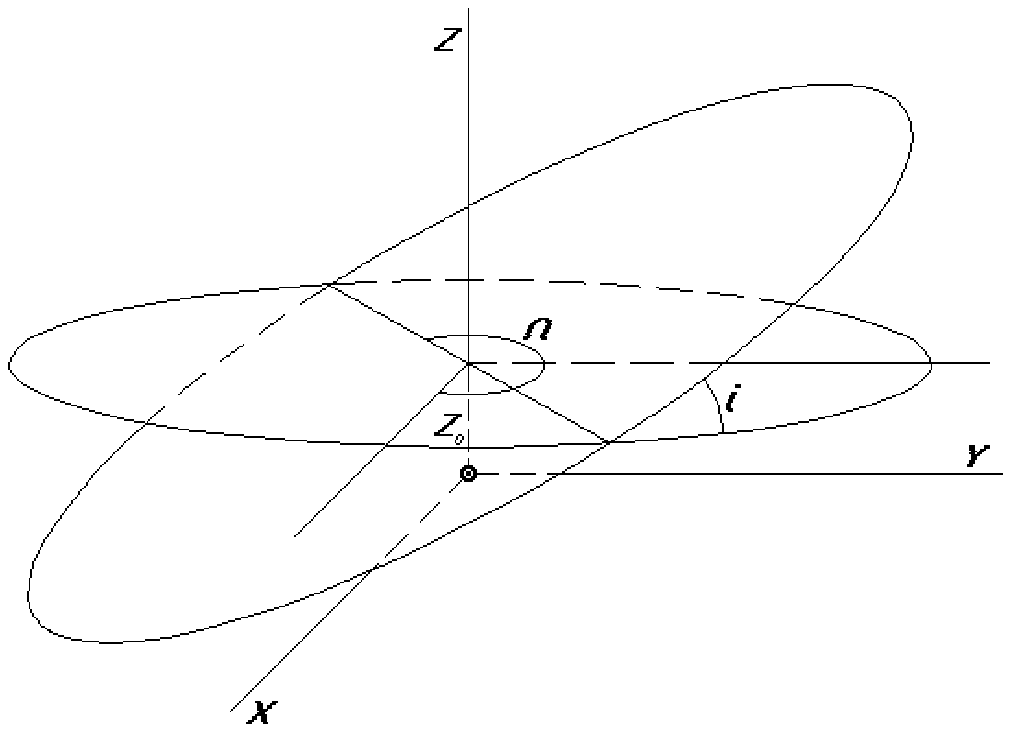} \caption{Geometrical scheme of the
two intersecting planes model. See the text for explanation of the
symbols.}
\end{figure}

\subsection{Statistical procedure}

The classification of the star sample will be done according to the principles
of Bayesian discriminant analysis, a description of which may be
found in \citet{c1}.

If we know all the parameters that define the planes, we can
construct the pdfs for the GB and the LGD, $\phi_{GB}$ and
$\phi_{LGD}$, from equation (1). Then, if we know the a priori
probability of any star being a GB member, $f_{GB}$, the Bayes
theorem can be written as

\begin{equation}
p_{GB} = \frac{f_{GB}\phi_{GB}}{f_{GB}\phi_{GB} + (1-f_{GB})\phi_{LGD}}
\end{equation}

where $p_{GB}$ is the {\it a posteriori\/} probability of belonging to the
GB for a star with a known distance to both planes. Thus, following
the Bayes minimum error rate decision rule, we may classify a star
as a GB member if $p_{GB} > 0.5$, or as a LGD member if $p_{GB} \leq
0.5$.

Obviously, since we lack a pre-classified sample from which to
obtain the parameters of the planes, we must follow an iterative
procedure to estimate them. Departing from some reasonable initial
values of $i$, $\Omega$ and $Z_0$ for the planes of the GB and the
LGD, of $f_{GB}$ and of the scale heights $h_{GB}$ and $h_{LGD}$,
that we use for the initial classification of the sample, we obtain
a first result that serves to calculate a second estimate of all the
parameters, which we then use as the initial values for a third
estimation; and so on, iteratively, until we reach the convergence.

\subsubsection{Main algorithm}
The estimation algorithm is divided into the following steps:

\begin{enumerate}
\item The scale heights $h_{GB}$ and $h_{LGD}$ as parameters of the pdfs
are simply estimated as the variance of the distances of the stars to the corresponding
plane.
\item $f_{GB}$ is estimated as the ratio of the obtained number of GB members to  the sample size.
\item The parameters of the planes are obtained by orthogonal least
squares fitting. If ${\mathbf r}_i=(x_i,y_i,z_i)$ is the position
vector of the star $i$, we first calculate the mean position vector
$\overline{{\mathbf r}}$ and the matrix of moments ${\mathbf M}$

\begin{equation}
\overline{{\mathbf r}}=\frac{1}{N}\sum_{i=1}^{N}{\mathbf r}_i\\ \\
\end{equation}

\begin{equation}
{\mathbf M}=\frac{1}{N}\sum_{i=1}^{N}({\mathbf r}_i -
\overline{{\mathbf r}})^T({\mathbf r}_i - \overline{{\mathbf r}})
\end{equation}

where $N$ is the number of stars in the classified sample. We solve
the eigenvalues equation

\begin{equation}
\det({\mathbf M}-\lambda {\mathbf I})=0
\end{equation}

and then we solve for the eigenvector ${\mathbf u}$

\begin{equation}
({\mathbf M}-\lambda_0 {\mathbf I}){\mathbf u}=0
\end{equation}

where $\lambda_0$ is the smallest root of equation (8). Then the
fitted orthogonal least squares plane equation may be written

\begin{equation}
{\mathbf u} \cdot ({\mathbf r} - \overline{{\mathbf r}})=0
\end{equation}

and from the values of the $a_i$ coefficients, we can compute $i$,
$\Omega$ and $Z_0$, according to equation (4).
\end{enumerate}

This iterative algorithm is fully nested within a bootstrap
structure in order to simultaneously obtain error estimates of the
model parameters. Thus, the procedure is performed once with the
true sample, and then it is repeated 99 more times with {\it
pseudo-samples\/} of the same size built by choosing stars at random
-repetition being allowed- from the original batch.

\subsubsection{Detection of outliers}
An important issue remains, though, which is that of the possible
contamination of our sample by outliers that don't belong to either
of our two distributions as we have defined them, i.e., that they
are too far away from the mean planes and cause deviations in the
estimation of the parameters. This implies that an outlier will be
found in zones of low density of probability in the sampling space \citep{c0}.
Since we know the pdfs for the GB and the LGD, we can evaluate the
total density $D$ in the position of any star of the original
sample:

\begin{equation}
D = f_{GB}\phi_{GB} + (1-f_{GB})\phi_{LGD}
\end{equation}

We consider as an outlier any star located in a point in space
where the density D is lower than a certain threshold, which
indicates the regions so far from the mid planes that we may
safely consider that stars in these zones do not belong to
our distributions. We have found after the examination of our system
that a conservative value of $D < 10^{-3}$ fulfills our requirements
to eliminate most of the possible outliers. All the outliers are removed
from the sample before the iterative process begins and the new parameters are
estimated. Once the new result has been obtained, we reintroduce the
full sample and obtain the outliers as defined  by the new planes,
thus beginning a second iterative procedure that will lead to the
final estimation of the parameters without contamination by
outliers.

\subsection{Stability analysis}
In order to study the behavior of our model, we have built a series
of test samples with known parameters that we may compare with those
estimated by the algorithm. These samples respond to the
simplifications assumed in our model, i.e. they are two clouds of
points distributed along two crossed mid planes. In rectangular
coordinates ($X$,$Y$,$Z$), the points are originally distributed at
random in the $XY$ plane, while their height $Z$ follows an exponential pdf (equation 1).
Then, each cloud of points is rotated as a
whole by an angle $i$, and then by an angle $\Omega$, and displaced
a vertical distance $Z_0$ (their values being obviously different
for each set of points). Finally, a distance limit $r$ to the whole
sample is chosen ($\sqrt{a_1^2 + a_2^2 +a_3^2} \leq r$). The number of points
in each sample is about 700 to 800, which are the remaining "stars"
after the distance cut (each sample originally has 1000 points which,
to avoid undesired border effects, extend beyond the 1 kpc limit when we assign
their random distances). A typical test sample after the separation procedure
is shown in Figure 2, where 5\% of the LGD stars and 7.5\% of the GB stars have been
misclassified, mainly in the zone of intersection between the
planes. Some isolated stars far away from the mid plane of their
corresponding distribution have also been misclassified.

\begin{figure}[h]
\epsscale{.90} \plotone{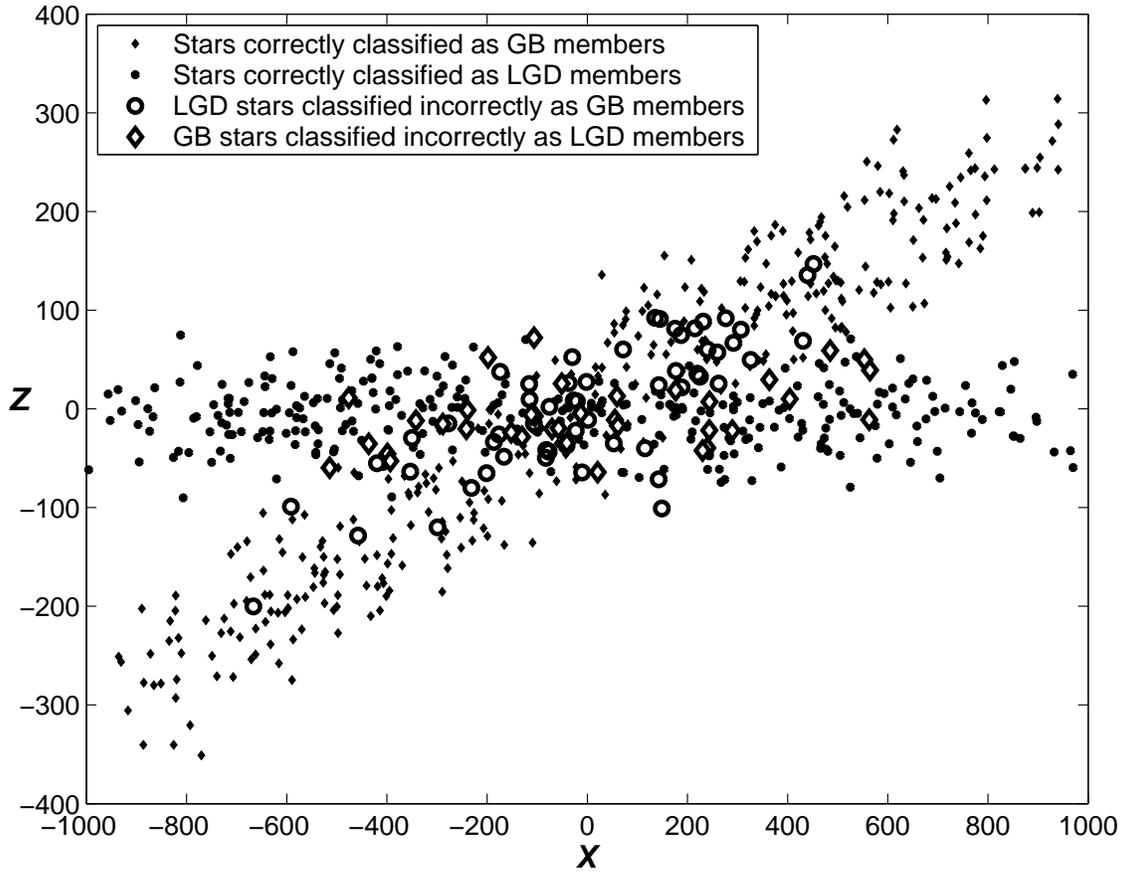} \caption{Classification of a
synthetic sample of stars with parameters $i_{GB}=17\degr$,
$\Omega_{GB}=285\degr$ and $Z_{0}^{GB}=-10$}
\end{figure}

In Figure 3 we plot the results of one series of tests of a
simulated GB for each of the model parameters. Each point in the top panel of Figure 3
represents the mean result of the first iteration procedure for a hundred samples with
fixed true parameters ($i_{GB} = 17\degr$, the dashed line in the figure). The initial
values introduced at the beginning of the process are equal to the true parameters of
the synthetic samples, except for the one whose behavior we want to analyze; in
this case, $i_{GB}$ . For each of these points we observe that:
\begin{enumerate}
  \item Convergence is reached typically in less than 10 iterations, for a precision of $0.01\degr$.
  \item The mean error of $i_{GB}$, estimated by the bootstrap procedure, is $0.2\degr$, so the
  number of iterations actually needed for a significative convergence is very low.
\end{enumerate}

We also observe that the first iterative process leads to an underestimation of the
inclination if the initial value is lower than the real one, or to an overestimation
if the initial value is greater the true inclination. A balance is achieved by the
reintroduction of the full sample for a second iteration under the newly estimated
parameters, which leads to a new, more refined detection of the outliers. For instance,
for an initial value of $i_{GB} = 14\degr$, an initial convergence is reached at
$i_{GB} = 16.4\degr$; introducing $i_{GB} = 16.4\degr$ as the initial value leads to a
final convergence of $i_{GB} = 16.8\degr$. Only two steps have been
necessary to stabilize the result, this being the typical behavior
in all the tests performed.

\begin{figure*}
\epsscale{1.0} \plotone{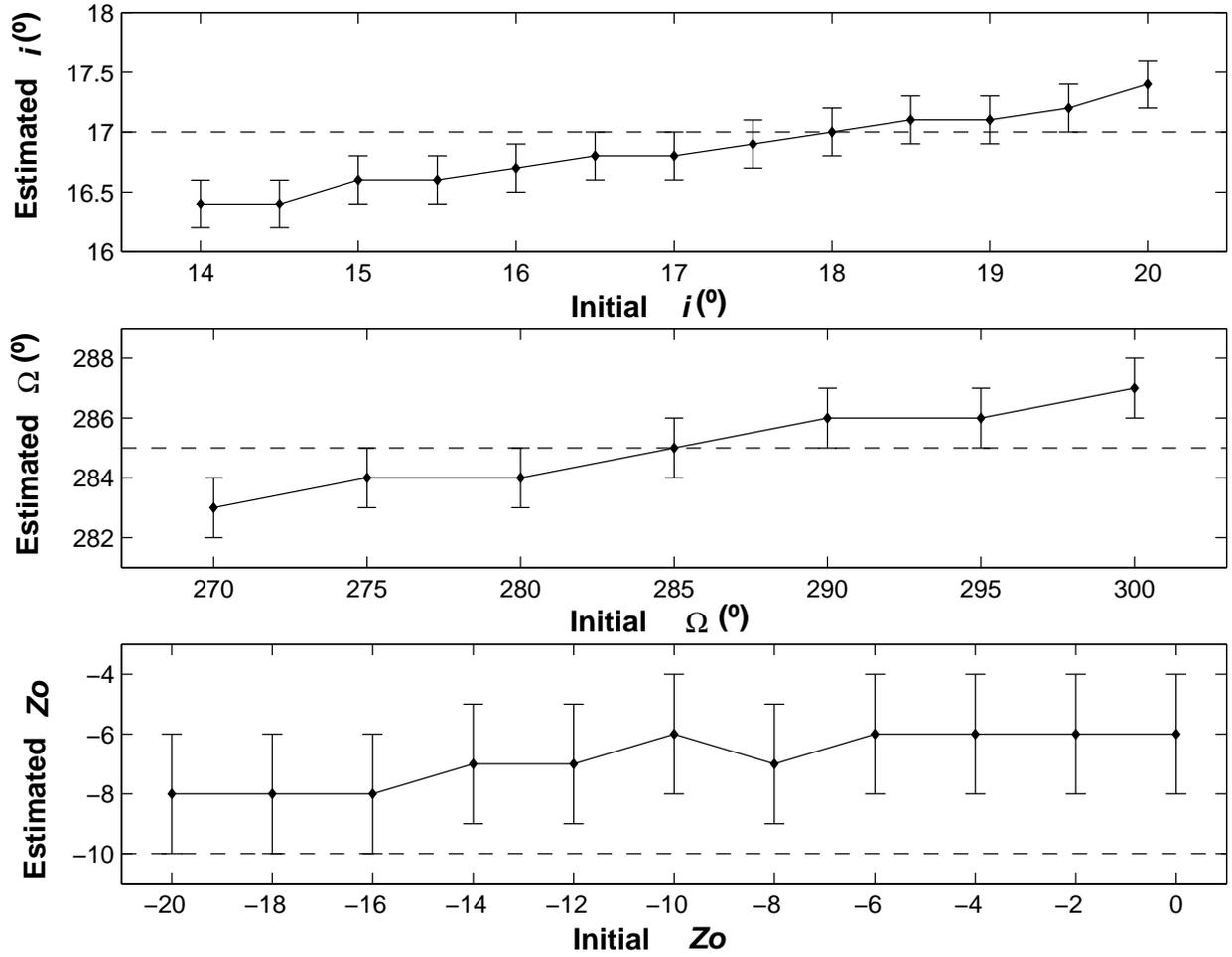} \caption{Estimation tests for a
synthetic GB with the same parameter values as in Figure 1 (marked
here by dashed lines). Each point represents the mean estimation of
100 tests with the correspondent initial value}
\end{figure*}

\begin{figure*}
\epsscale{1.0} \plotone{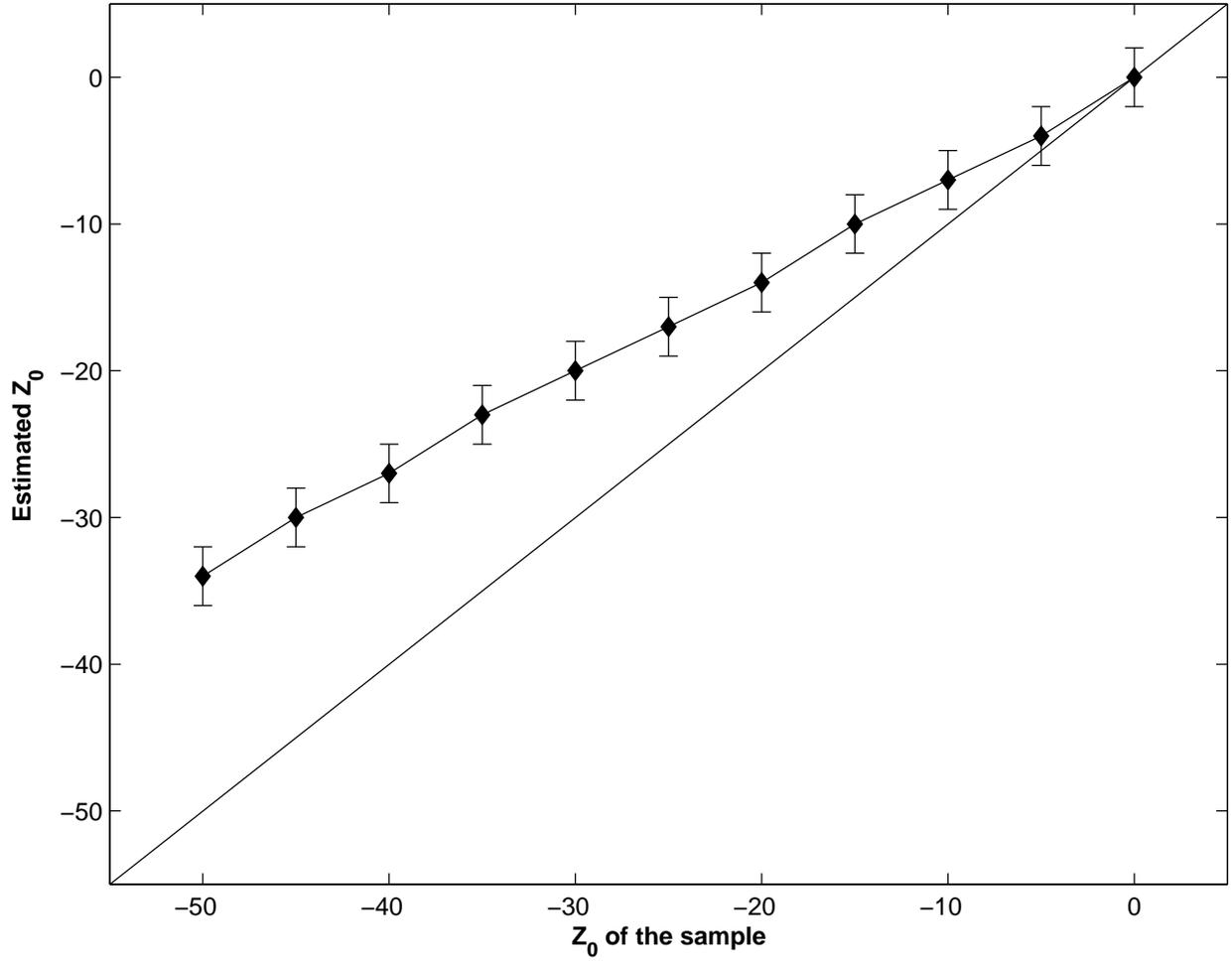} \caption{Estimation tests for a
synthetic GB with different values of $Z_{0}$. Each point represents
the mean estimation of 100 tests with the correspondent true value
of the sample}
\end{figure*}

Similar tests for $\Omega$ and $Z_0$ are shown in the middle and
bottom panels of Figure 3, respectively. While the results for the
longitude of the ascending node are even more precise than those
obtained for the inclination (convergence is reached at $\Omega_{GB}
= 285\degr \pm 1\degr$ for a true value of $\Omega_{GB} =
285\degr$), an underestimation -in absolute value- of the vertical
distance to the Sun is observed (convergence is now reached at
$Z_{0}^{GB} = -6 \pm 2$ pc for a true value of $Z_{0}^{GB} = -10$
pc). This being the parameter of the plane the most sensible to the
dispersion of the stars, it seems unavoidable that we must cope with
a certain bias in the estimation of $Z_0$. In order to evaluate the
magnitude of this systematic difference we have performed some
additional tests, whose results are shown in Figure 4. A hundred
aleatory samples have been built for each value of $Z_{0}^{GB}$ from
0 to 50, then we have run the program to estimate $Z_{0}^{GB}$ and
its error for each case. We represent their mean values in the
figure, from which we can see how the estimation of $Z_{0}^{GB}$ is
affected by a systematic bias of around $30\%$.

We must note that similar tests performed systematically for
$Z_{0}^{LGD}$ show that there is no such a bias in the case of the
LGD (or, at least, it is smaller than the random errors in the
estimation), probably due to its negligible inclination. This points
out that it is the geometry of the problem which is responsible for
such behavior of our algorithm, and thus any pdf that we consider
as a model for the vertical distribution of the planes does not
alter this bias. We have corroborated that this is true by
simulating the LGD and the GB systems with a Gaussian vertical
distribution of the stars, and then performing the same tests with a
Gaussian pdf model in our program. The results are consistent with
the ones obtained working with exponential distributions, i.e., the
LGD shows no significative bias in $Z_{0}^{LGD}$, and $Z_{0}^{GB}$
is affected by a systematic bias of around $35\%$.

When the simultaneous convergence of all the parameters ($i$,
$\Omega$, $Z_0$ and $h$ for both planes, and $f_{GB}$) is sought,
again only about ten iterations for the first loop and two or three
for the second loop are needed, thus confirming the stability of
the model.

We want to remark the importance and the uniqueness of the detection of outliers
in our procedure. Stars located in the extremes of the sample distribution may
have a great weight in the estimation of the model parameters, but according to
their low probability of belonging to the model distribution, they must be considered
as probable outliers and thus eliminated from the process. This purge of outliers leads
to quite robust results, which is essential when dealing with a system such as the GB,
whose spatial distribution cannot be delimited without a certain degree of uncertainty.

\section{Star sample}

We have selected a star sample from the {\it Hipparcos\/} Catalog
\citep{e4} with spectral types from O to B6 and luminosity classes
III, IV and V. The spectral types listed in the {\it Hipparcos\/}
Catalog are a compilation from different sources; thus, a lack of
homogeneity in the precision of the spectral type or the luminosity
class may be present in the data. There are not systematic studies
devoted to analyze the reliability of the spectral information
contained in the catalog. The closest work to such an analysis has
been performed by \citet{a0}, who obtained and classified spectra
for 584 stars belonging to {\it A supplement to the bright star
catalog\/}. The comparison between this classification and that in
the {\it Hipparcos\/} Catalog yields that the estimated error in the
spectral classification is $\pm 1.2$ subtypes, and that a $10\%$ of
the listed luminosity classes may be wrong. In this paper, the
spectral classification is only used as a rough estimate of the
stellar ages. Thus, the uncertainties in the spectral classification
will not have any influence in the main results of our study.

For each star, the following data were chosen:

\begin{itemize}
\item HIP, {\it Hipparcos\/} identifier number.
\item Trigonometric parallax $\pi$ (mas)
\item Standard error in trigonometric parallax, $\sigma_{\pi}$ (mas)
\item Right ascension for the epoch J1991.5 in IRCS (degrees)
\item Declination for the epoch J1991.5 in IRCS (degrees)
\end{itemize}

Also, $uvby\beta$ Str\"omgren photometry data from the  catalog
of \citet{h2} were used to calculate the photometric distance of
every star in the sample through the \citet{b2} calibration.

Since the relative error ($\sigma_{\pi}/\pi$) of the distances
estimated by {\it Hipparcos\/} parallaxes grows with the
trigonometric distance ($1/\pi$), we have decided to keep the
trigonometric distance for a star only if the relative error in its
parallax is lower or equal to a $10\%$ (which corresponds
approximately to distances closer than 100 pc from the Sun);
otherwise, the photometric distance is chosen. Not only the error in
the Str\"omgren estimation is independent of the distance, but also
we can see (Figure 5) that the distribution of photometric distances
is very similar to the distribution of trigonometric ones, and that
no systematic trends can be found. Comparison between the medians of
both distance estimations for stars with a relative error in their
parallaxes lower or equal to a $20\%$ yields a difference between
them of merely $1\%$. This is in good agreement with the studies
performed by \citet{k1}, who find no significant difference between
{\it Hipparcos\/} trigonometric distances and those obtained from
$uvby$ and $H_{\beta}$ photometry. Thus, we rest assured that the
use of Str\"omgren photometric distances will not harm the precision
of our three-dimensional picture of the solar neighborhood.

\begin{figure}[h]
\epsscale{0.8} \plotone{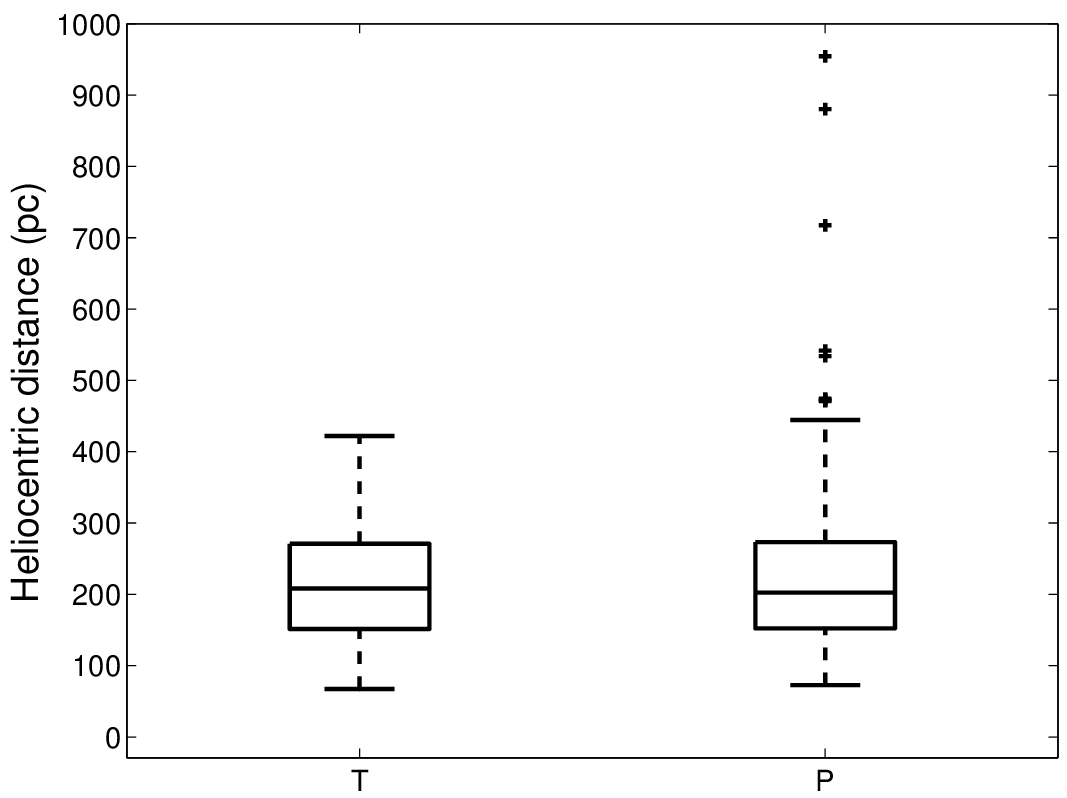} \caption{Boxplot comparison between
\textit{Hipparcos} trigonometric distances (T) and photometric
distances in the Str\"omgren system (P) for stars with a relative
error in the parallax lower or equal to a $20\%$}
\end{figure}

As we said, part of our sample has been selected with a relative
error in its parallax lower or equal to a $10\%$. While the distance
bias due to the non-linear relationship between parallax and
distance is negligible if $\sigma_{\pi}/\pi$ is smaller than a
$10\%$ \citep{a1}, the Lutz-Kelker bias \citep{l5} caused by
truncating a sample based on the observed parallax relative error is
more difficult to evaluate and depends on the parent population, as
well as on the size of the selected subsample. In our case only 28
stars (about a $5\%$ of the total sample) have been selected by the
parallax criterium. Thus, although this part of the sample may be
affected by the Lutz-Kelker bias, the total sample shows little
contamination by this effect, so wa can assume that this truncation
bias is negligible in our final selection.

Finally, we have eliminated the stars with an estimated distance
greater than 1 kpc, because, according to results found in the literature,
the typical maximum radius of the GB is not greater than 700 pc \citep{s1,w1,c4,f1}.
The remaining sample is composed of 553 stars; Figure 6 represents their spatial
distribution projections in the three Cartesian planes $XY$ (top panel), $XZ$
(middle panel) and $YZ$ (bottom panel).

\begin{figure*}
\epsscale{0.8} \plotone{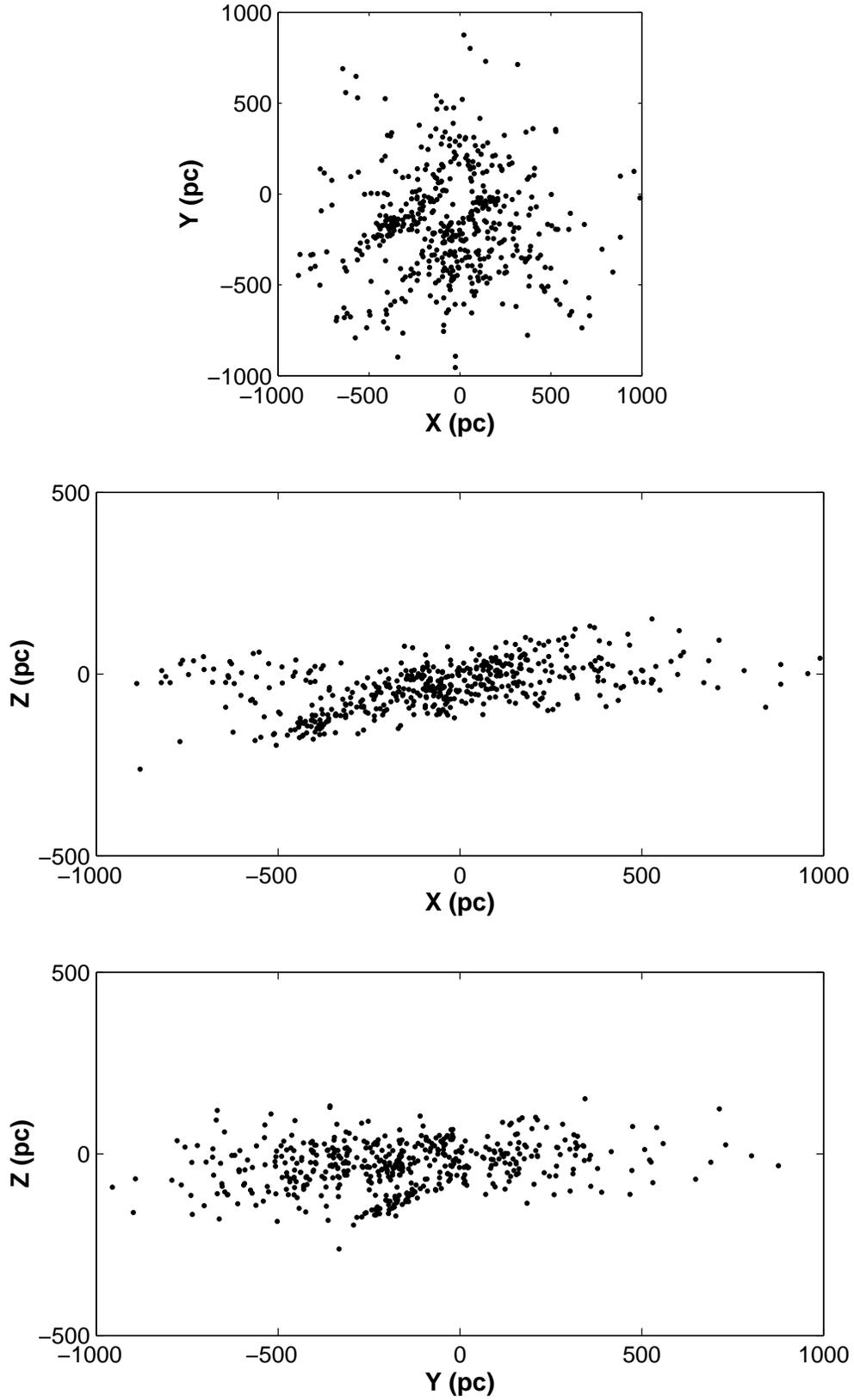} \caption{Spatial projections of the
star sample}
\end{figure*}

\section{Structural results}

We apply our model to the total star sample, as well as to four cumulative
sub-samples that comprise all the spectral types from O to B2, B3,
B4 and B5, respectively. In all cases, we find that the distribution
of the stars can be satisfactorily described by our model of two intersecting
planar systems, the GB presenting a remarkable inclination with respect to the LGD.
While the stars belonging to the latter tend to be homogeneously distributed
on the $XY$ plane, the former shows a more clumpy, filamentary structure
(yet the global inclination is maintained across the whole GB system).
The numerical results that fit the model planes for the cumulative samples
of increasing spectral types are displayed on table~\ref{tbl-1}. We have listed separately
the solutions depending on whether an exponential or a Gaussian pdf has been used.

While some of the model parameters are found to be the same -within
the estimated errors- regardless of the employed pdf, others show
non negligible differences. Specially, the number of outliers and
the fraction of stars belonging to the GB system ($f_{GB}$) display
significant discrepancies. The number of outliers working with a
Gaussian pdf is close to a $20\%$ of the total sample, meaning that
about 100 out of 554 stars have a very low probability of belonging
either to the GB or to the LGD. This fraction, though, is reduced to
a $10\%$ when an exponential pdf is used. According to these results
we consider that our sample is better described by an exponential
law than by a Gaussian one. We must note that this cannot be
extrapolated for the GB system as a whole, and thus must be seen
only as the best fitting for our observational sample.

\begin{deluxetable}{lcccccccccccc}
\tablecolumns{13}
\tablewidth{0pt}
\tabletypesize{\scriptsize}
\tablecaption{Estimated parameters of the GB and the LGD planes\label{tbl-1}}
\tablehead{
\colhead{} & \colhead{} & \colhead{} & \colhead{} &
\colhead{} & \colhead{$h_{GB}$} & \colhead{$h_{LGD}$} &
\colhead{$i_{GB}$} & \colhead{$i_{LGD}$} &
\colhead{$\Omega_{GB}$} & \colhead{$\Omega_{LGD}$} &
\colhead{$Z_{0}^{GB}$} & \colhead{$Z_{0}^{LGD}$} \\
\colhead{$Sp$} & \colhead{$N$} & \colhead{$N'$} & \colhead{$n$} &
\colhead{$f_{GB}$} & \colhead{$(pc)$} & \colhead{$(pc)$} &
\colhead{$(\degr)$} & \colhead{$(\degr)$} & \colhead{$(\degr)$} &
\colhead{$(\degr)$} & \colhead{$(pc)$} & \colhead{$(pc)$} }
\startdata
\sidehead{Exponential pdf:}
O-B2  &  181 & 162 & 3 &  0.53 (0.06) & 27 (4) & 34 (3) & 16 (2) & 2 (1) &  273 (7) & 354 (165) &  -14 (8)  & -10 (8)\\
O-B3  &  301 & 267 & 2 &  0.54 (0.05) & 27 (3) & 34 (2) & 14 (1) & 2 (1) &  278 (5) & 352 (152) &  -19 (10) & -10 (8)\\
O-B4  &  341 & 303 & 3 &  0.54 (0.05) & 27 (3) & 35 (2) & 14 (1) & 2 (1) &  281 (5) & 354 (154) &  -17 (9)  & -17 (7)\\
O-B5  &  484 & 433 & 1 &  0.54 (0.05) & 27 (3) & 35 (2) & 14 (1) & 1 (1) &  284 (3) & 356 (156) &  -13 (5)  & -17 (5)\\
O-B6  &  553 & 498 & 1 &  0.54 (0.05) & 27 (3) & 34 (2) & 14 (1) & 1 (1) &  284 (3) & 355 (149) &  -13 (6)  & -16 (5)\\
\sidehead{Gaussian pdf:}
O-B2  &  181 & 144 & 3 &  0.54 (0.06) & 18 (3) & 23 (3) & 17 (1)   & 1 (1) &  277 (5) & 346 (147) &  -4 (5)  & -10 (5)\\
O-B3  &  301 & 247 & 5 &  0.49 (0.05) & 18 (2) & 27 (2) & 16 (1)   & 1 (1) &  280 (3) & 342 (126) &  -6 (5)  & -12 (4)\\
O-B4  &  341 & 279 & 2 &  0.48 (0.06) & 18 (2) & 28 (2) & 16 (1)   & 1 (1) &  280 (4) & 342 (116) &  -7 (3)  & -14 (4)\\
O-B5  &  484 & 396 & 4 &  0.40 (0.04) & 15 (1) & 30 (2) & 17 (0.4) & 1 (1) &  282 (2) & 327  (78) &  -8 (3)  & -15 (3)\\
O-B6  &  553 & 454 & 4 &  0.37 (0.03) & 13 (1) & 30 (2) & 17 (0.3) & 1 (1) &  284 (2) & 321  (58) &  -5 (3)  & -16 (3)\\
\enddata
\tablecomments{$Sp$ stands for the spectral types comprised in each
cumulative sample, $N$ gives the initial sample size, $N'$ is the
number of stars remaining in the sample after the elimination of
outliers and $n$ is the number of iterations in the second loop
needed to reach convergence (it must be noted that the initial
values of the parameters given to the algorithm are the ones
obtained as a result from the previous subsample, except in the O-B2
case). It must be also noted that $h$ stands for both the scale
height of the exponential model, and for the half-width of the
Gaussian pdf solution. Errors estimated by bootstrap are given in
parentheses.}
\end{deluxetable}

We find an inclination of the GB between $14\degr \pm 1\degr$ and
$17\degr \pm 0.3\degr$ for all the samples, except for the youngest
sub-sample of O-B2 stars, where the range is reduced to $i_{GB} =
16\degr \pm 2\degr - 17\degr \pm 1\degr$ ). It's a smaller value
than those commonly found in the literature, yet we may compare it
with those estimated by \citet{w1} on table~\ref{tbl-2}. In his
study he estimates too a larger inclination for the youngest sample
(albeit $2\degr$ larger than that found for our sub-sample with the
earliest spectral types). Moreover, his result for the sample
ranging from 30 to 60 Myr matches the inclination of $14\degr$ found
for our later spectral types. Also, \citet{t2,t3}, from a sample of
{\it Hipparcos\/} OB stars with Str\"omgren photometric distances,
estimate an inclination of $22\degr$ for stars younger than 30 Myr,
and an inclination of $16\degr$ for stars in the range of 30 to 60
Myr. It's worth noting that in the earlier work of \citep{s1},
$i_{GB} = 18\degr \pm 0.4\degr$ for their O-B5 sample, and $i_{GB} =
19\degr \pm 1\degr$ for an O-B2.5 subsample, yet they find for stars
in a range of spectral types from B3 to B5 that $i_{GB} = 16\degr
\pm 1\degr$.

\begin{deluxetable}{lcc}
\tablecolumns{3}
\tablewidth{0pt}
\tabletypesize{\scriptsize}
\tablecaption{GB plane parameters for samples of different age, from
\citet{w1}\label{tbl-2}}
\tablehead{ \colhead{Sample Age} &
\colhead{$i_{GB}$} & \colhead{$\Omega_{GB}$} \\
\colhead{$(yr)$} & \colhead{$(\degr)$} & \colhead{$(\degr)$}}
\startdata
$< 2 \cdot 10^{7}$    &  $18.9 \pm 1.1$ & $273.2 \pm 3.1$ \\
$< 3 \cdot 10^{7}$    &  $18.1 \pm 0.9$ & $270.9 \pm 4.3$ \\
$(3-6) \cdot 10^{7}$  &  $14.3 \pm 2.0$ & $286.2 \pm 5.7$ \\
\enddata
\end{deluxetable}

Comparing the values that we estimate for $\Omega_{GB}$, we see that
they are in very good agreement with those of \citet{w1}. The
earlier spectral types present a smaller value of $\Omega_{GB}$,
similar to that found by \citet{w1} for the youngest samples, while
his result for the oldest age groups matches those we obtain when
later spectral types are included. The values of $\Omega_{GB} =
275\degr - 295\degr$ (depending on the sample age) found by
\citet{t2,t3} are also in good agreement with our results. Other
similar values were estimated by \citet{t1}, $\Omega_{GB} = 278\degr
- 290\degr$, and \citet{c4}, $\Omega_{GB} = 284.5\degr$, from OB
stars of the {\it Hipparcos\/} proposal.

\begin{figure*}
\epsscale{1.0} \plotone{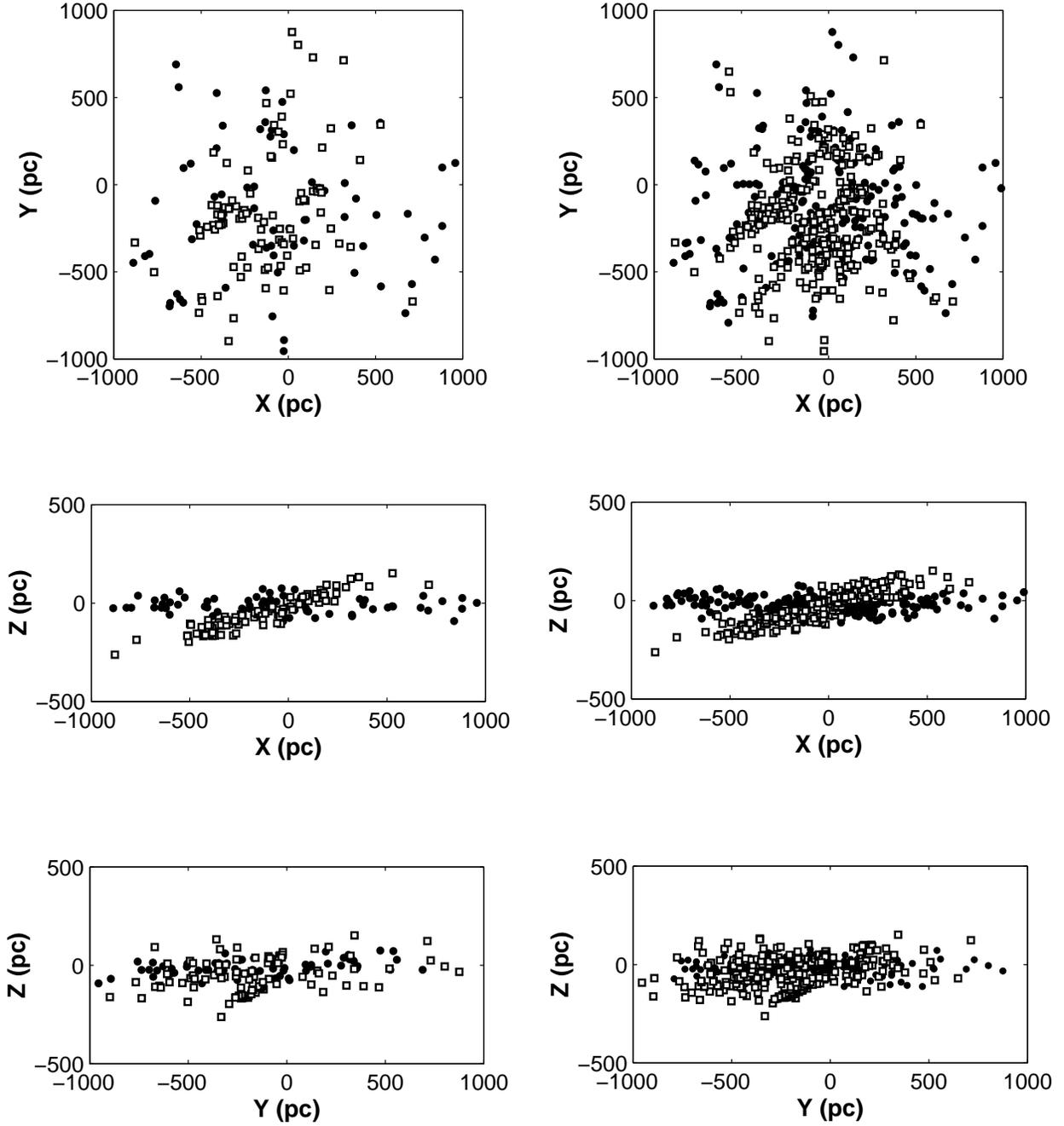} \caption{Classified O-B2 sub-sample
(left) and full O-B6 sample (right). Black circles represent the LGD
and open squares, the GB.}
\end{figure*}

Even though there's agreement in most of the estimations made by
different authors of the GB geometrical parameters around a certain
range of values, it is true that this range is sometimes quite large
(for instance, $i_{GB} \approx 14\degr$ - $22\degr$).  We shall
venture that this isn't only caused by the choice of different ages,
spectral types or heliocentric distance limits of the star samples,
but that the presence of outliers in those samples may seriously be
affecting the results. It is perfectly possible that the weight of
only a few stars with a low membership probability to the distribution
sensibly alters the estimation, leading to unrealistic values of the
parameters. Since the "limits" of the GB constitute a very diffuse
zone whose boundaries cannot be easily defined, we consider that
unless outliers are purged it is not possible to characterize its
spatial distribution with confidence enough as to define the series of
parameters (such as the inclination or the longitude of the ascending
node) that are usually employed to describe its geometry.

Our model also gives an estimation of the Sun's distance to the
Galactic plane, which oscillates between $10 \pm 8$ pc (for the O-B2
and O-B3 subsamples) and $17 \pm 7 / 5$ pc (for the O-B4 and O-B5
subsamples, respectively). Although this parameter is biased, as we had
already foreseen in the simulations testing the model (the bottom panel of Figure 3, and
Figure 4, perfectly illustrate that), the raw values we obtain are compatible with
those found by \citet{h4} from the Palomar Sky Survey star counts
($Z_{\odot} = 20.5 \pm 3.5$ pc), by \citet{h1} from $2.2 \mu m$ and
$3.5 \mu m$ maps by the DIRBE instrument of the COBE satellite and the
Two-Micron Galactic Survey ($Z_{\odot} = 15.5 \pm 3$ pc), or by
\citet{c3} from the IRAS Point Source Catalog in $12 \mu m$ and $25
\mu m$ ($Z_{\odot} = 15.5 \pm 0.7$ pc). However, the mean age of these catalogs
is much older than the young LGD we are dealing with in this work.


The comparisons of the O-B5 subsample with the
analysis of a star sample of the same spectral types by \citet{s1},
who choose an exponential-type scale height model, show an even
better match with our results. They find that $Z_{\odot} = 24 \pm 3$
pc and that -in the sphere of 200 pc around the Sun- the scale
height of the LGD is $h_{LGD} = 45 \pm 18$ pc. Their estimation of
the scale height of the GB, $h_{GB} = 27 \pm 4$ for stars closer
than 200 pc and $h_{GB} = 27 \pm 1$ for stars closer than 800 pc, is
in perfect agreement with our results. A more recent study on the
disk O-B5 stellar population by \citet{m0}, using a highly developed
Gaussian model, concludes that $Z_{\odot} = 25.2 \pm 2.0$ pc and
$h_{LGD} = 62.8 \pm 6.4$ pc. Also, working with a self-gravitating
isothermal disk model, he finds that $Z_{\odot} = 24.2 \pm 2.1$ pc
and $h_{LGD} = 34.2 \pm 3.3$ pc, the latter value matching almost
exactly our estimation of $h_{LGD} = 35 \pm 2$ for the O-B5
subsample with an exponential model. It is worth noting too that our
earlier O-B2 subsample yields similar results to those obtained by
\citet{r1,r2} for an O-B2 disk using also an exponential model
($h_{LGD} = 45 \pm 20$ pc, $Z_{\odot} = 9.5 \pm 3.5$ pc).

We also estimate a small inclination for the LGD between $1\degr \pm 1\degr$
and $2\degr \pm 1\degr$, which -although its high uncertainty points out that
it isn't a very significative value- is somewhat greater than that by \citet{h1}
of $0\degr.40 \pm 0\degr.03$. It is small enough, though, to make $\Omega_{LGD}$
remain not well defined, hence the large ($150\degr - 160\degr$)
uncertainty that we should expect for pure geometrical reasons
when $i_{LGD}$ is close to zero, as it is the case.

Figure 7 displays the three possible projections for the classified
O-B2 sub-sample (left panels) and the complete O-B6 sample (right panels) for the
exponential pdf solution. The inclination of the GB is clearly seen in the $XZ$ plane (middle panels);
it is also interesting to observe the irregular spatial distribution of
the GB in the $XY$ plane (top panels) presenting clustering and
filamentary structures, in contrast with the more homogeneous LGD distribution.
That is better seen in Figure 8, where the spatial density distribution of both systems has been
plotted. Note the clustering of the GB density distribution (left panel), the
most prominent feature being located in the Orion region in the
third Galactic quadrant.

\begin{figure*}
\begin{center}
\includegraphics[width=18cm,angle=0,scale=0.75]{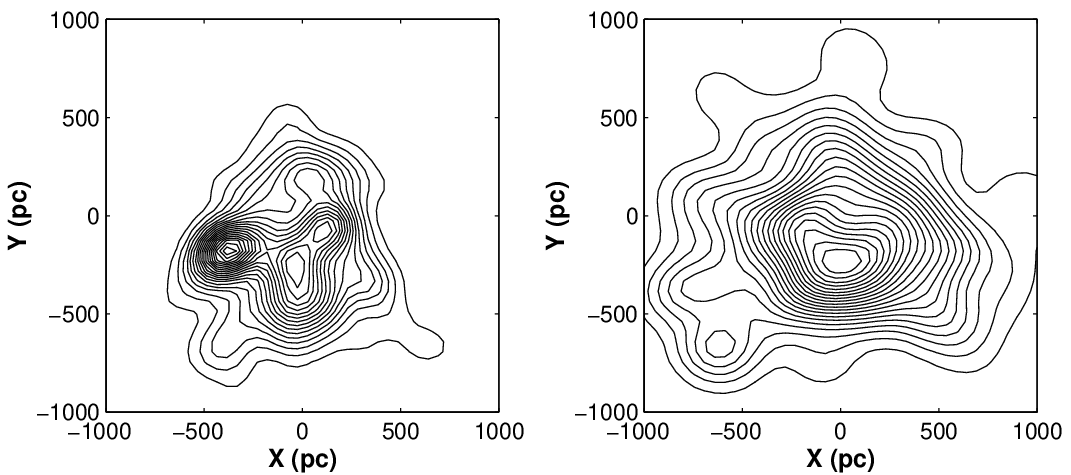}
\caption{Spatial star density of the
GB (left) and the LGD (right) for the classified full sample.}
\end{center}
\end{figure*}

\section{Correction of completeness}

The spatial density distribution of the LGD shows that the
star density decreases with the increasing distance to the Sun,
as we may see in the right panel of Figure 8. Certainly, such a
distribution is not what we should expect on this region of
the solar neighborhood. If we assume that on this scale the LGD
should present a homogeneous star density, it is evident that what
we observe in the figure is due to the incompleteness of our sample.
Thus, under this hyphotesis, we can use the two-dimensional density of
the LGD as a measure of the star field incompleteness in the working
region. This density map may be seen as a sort of "flat field"
used for completeness correction, that will allow us to improve the
structural analysis of the GB.

We have to note, though, that the interstellar medium extinction pattern
may also be playing an important role in the modulation of the observed density
structure. While the effect of distance on the incompleteness of the sample is
indistinctly reflected on the density correction for the GB and the LGD, the
extinction pattern could show different features for both systems. Thus, this
"flat field" correction must be considered as a first approach to the problem of
incompleteness, and not as a definitive solution. However, the main properties
of the dust distribution in the solar neighborhood (see P\"oppel 1997 for a detailed
discussion) makes us confident in the reliability of this approach.

\subsection{Model parameters}
Dividing the GB star density by the LGD star density would be enough
to provide a relative density that would enhance the structures that
belong solely to the GB, but we can take a further step and
introduce this correction in our model. We thus modulate the weight
of any star (which originally was the probability of belonging to
the correspondent plane) by a {\it completeness function\/} that
simply is the inverse of the LGD star density in the $(x,y)$
position of the star, evaluated by a Gaussian kernel.

Following again the iterative process for the O-B6 full sample with an
exponential pdf model, we reach convergence in the following values:

\begin{itemize}
\item $f_{GB} = 0.58 \pm 0.06$

\item $h_{GB} = 31 \pm 4$ pc

\item $h_{LGD} = 34\pm 5$ pc

\item $i_{GB} = 14\degr \pm 1\degr$

\item $i_{LGD} = 2\degr \pm 2\degr$

\item $\Omega_{GB} = 287\degr \pm 6\degr$

\item $\Omega_{LGD} = 352\degr \pm 28\degr$

\item $Z_{0}^{GB} = -15 \pm 12$ pc

\item $Z_{0}^{LGD} = -12 \pm 12$ pc
\end{itemize}

We observe that, within the error margin, the results are similar to
those we had previously obtained (table~\ref{tbl-1}). This indicates
that even with an incomplete sample, the structural results
estimated by our model are still valid.

\subsection{Spatial structure}
In Figure 9 we draw the nearest OB associations over the star
density maps of the GB. As we see, the correction of
completeness removes the contribution to the central peak by the
effect of the decreasing star density with the heliocentric
distance, allowing the region around Scorpio to stand by itself.
Also, the Orion peak moves farther away, towards the position of Ori
OB1.

It is important to note that this is the first time that a
correction of completeness has been implemented in order to compensate
for the always incomplete samples constructed to study the spatial
distribution of the GB.  No significant difference can be found in the
estimation of the GB structural parameters, even though the new
density map of its spatial distribution (Figure 9, right panel) shows
important modifications of the original distribution (Figure 9, left
panel) in the $XY$ plane. This is probably due to the geometry of the
system, dominated by the GB's inclination with respect to the Galactic
plane (hence, the $Z$ vertical distribution playing a fundamental
role) and by the outer parts of the structure, better separated from
the LGD. Nonetheless, the completeness correction is tremendously
reassuring if we consider the difficulty of gathering a complete
enough sample of GB stars, and it lets us look confidently at our
results. It also provides a most interesting picture of the effects of
incompleteness on the clumpy $XY$ distribution of the GB, and its
correspondence with OB associations.

\begin{figure*}
\begin{center}
\includegraphics[width=9cm,angle=270,scale=0.75]{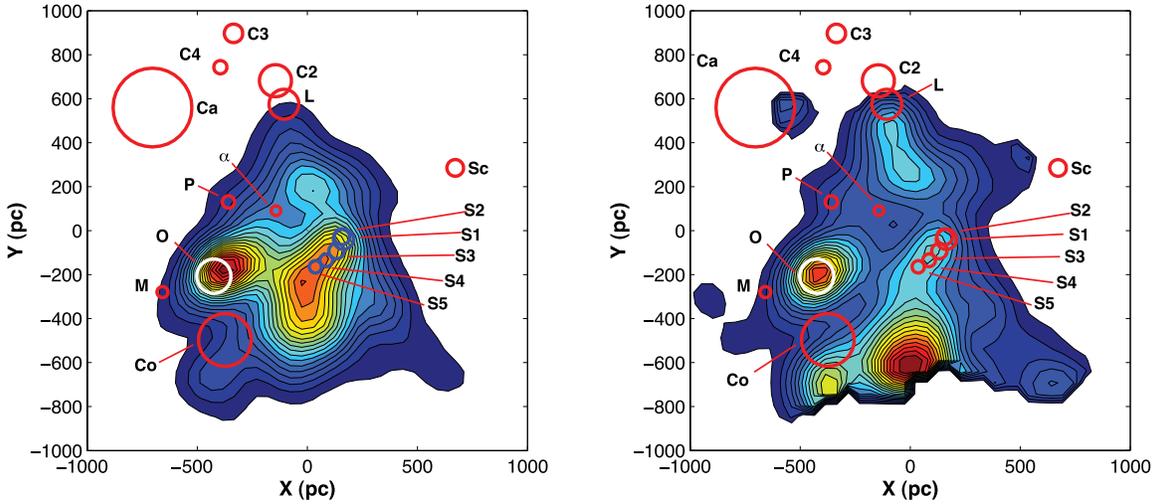}
\caption{Spatial star density of the GB before (left) and after
(right) the completeness correction. The circles represent the
pre-\textit{Hipparcos} OB associations, as listed by \citet{Zeu99}.
The letters stand for: Sco OB2$_{1}$ (S1), Sco OB2$_{2}$ (S2), Sco
OB2$_{3}$ (S3), Sco OB2$_{4}$ (S4), Sco OB2$_{5}$ (S5), Col 121
(Co), Ori OB1 (O), Mon OB1 (M), Per OB2 (P), $\alpha$ Per
($\alpha$), Cam OB1 (Ca), Lac OB1 (L), Cep OB2 (C2), Cep OB3 (C3),
Cep OB4 (C4), Sct OB2 (Sc)}
\end{center}
\end{figure*}

\subsection{Associations}

Since the work by Blaauw (1965), it is known that the young
associations Sco-Cen, Per OB2, and Ori OB1 are part of the GB, while
the field members could have originated from the disruption of oldest
groups (P\"oppel 1997). In fact, these three groups are well inside the
typical radius of the GB, but most of the associations found within 1
kpc in the radius around the Sun seem to be out of the classical
borders of the belt.

After the {\sl flat fielding} correction of the density map of the
GB, we observe how new clumps appeared (Fig. 9, right panel). Within
the uncertainty in the estimated distances of the associations,
these clumps can be linked to well known and cataloged associations
(e.g. de Zeeuw et al 1999). The GB membership of the associations is
not only defined by the position of the centroid (a single point),
but by a relative maximum in the stellar density pattern of the GB.
Thus, young associations as Cam OB1, Lac OB1, Col 121 and a clump
(at $X=0$, $Y\sim -600$ pc) that's connected to the Vela rift are
likely GB members, extending the GB frontiers to larger distances.

Perhaps the most striking result is the inclusion of some Vela groups as probable
GB members. We wonder -after a suggestion from the referee- if this presence
could be a spurious effect generated by the "flat field" correction due to a
different reddening pattern in the GB and the LGD. However, Vela is located close
to the line of nodes in which the GB and the LGD coexist. Thus, the dust distribution
is probably shared in this region by both systems, so it is not likely that it is
responsible for the introduction of this maximum.

\section{Conclusions}

We have developed a new three-dimensional spatial classification
method to estimate the structure of the GB and the LGD, working with
single star membership probabilities. This method, although based in
a model of two planar intersecting systems, allows for a greater
variety of spatial configurations than other classification
techniques found in literature, which is helpful in order to
understand the complex structure of the GB. We have tested this
method with artificial samples, and then we have applied it to a
true sample of OB stars in the sphere of 1 kpc around the Sun. As a
result, we have obtained that the distribution of young stars in the
solar neighborhood is dominated by two different systems: the LGD and
the inclined structure of the GB. Our hypothesis of two intersecting disks
has been quite effective in order to characterize both systems for their
study, and although we have found that the GB is a clumpy, filamentary
structure, we have successfully estimated through our model the geometrical
parameters of the GB and the LGD, which are in good agreement with earlier
estimations by other authors. The inclination of the GB is found to be
between $i_{GB} = 14\degr - 17\degr$.

Also, a completeness correction has demonstrated that the  estimation of the
global parameters that define our GB model were not affected by the decrease
of the star density with the heliocentric distance, providing  robust parameter estimators. Yet this correction
effectively refines the projected spatial density distribution of the GB
on the $XY$ plane, removing false substructures and enhancing the true ones,
some of which we may relate to the nearby OB associations.

\acknowledgments

We want to thank the referee, Jes\'us Ma\'{\i}z Apell\'aniz, for his
very sharp and useful comments that were very helpful in order to
improve the scientific content of this paper. F.E. wants to thank
the Departamento de F\'{\i}sica At\'omica, Molecular y Nuclear of
the Universidad de Sevilla for its support during this work. We
would also like to acknowledge the funding from MCEyD of Spain
through grants AYA2004-05395, and AYA2004-08260-C03-02, and from
Consejer\'{\i}a de Educaci\'on y Ciencia (Junta de Andaluc\'{\i}a)
through TIC-101.

\end{document}